\documentclass[aps,a4paper,prc, amsfonts, amssymb, amsmath, reprint, showkeys, nofootinbib, twoside, superscriptaddress, standalone]{revtex4-1}
\usepackage{lipsum}
\usepackage{newtxtext,newtxmath}
\usepackage{notoccite}
\usepackage{upgreek}
\usepackage{flushend}
\usepackage{array}
\usepackage{varwidth}
\usepackage[colorinlistoftodos, color=green!40, prependcaption]{todonotes}
\itshape
\usepackage{indentfirst}
\usepackage{pgfplots}
\usepackage{tikz}
\usetikzlibrary{shapes.geometric, arrows.meta, positioning}
\usepackage{multirow}
\graphicspath{{img/},{pics/}}	
\usepackage{braket}
\usepackage{graphicx}
\usepackage{float}
\usepackage{subfigure}
\usetikzlibrary{fit}
\usetikzlibrary{calc,positioning}
\usepackage{amsmath}
\usepackage{color}
\usepackage{hyperref}
\usepackage{calligra}
\hypersetup{hypertex=true,
	colorlinks=true,
	linkcolor=blue,
	urlcolor=blue,
	anchorcolor=blue,
	citecolor=blue}
\usepackage{longtable}	
\usepackage{supertabular}
\usepackage{titlesec}

\titleformat{\subsection}[hang]{\bfseries}{}{0.5em}{}
\titlespacing{\subsection}{0pt}{12pt}{8pt}

\usetikzlibrary{backgrounds}
\usetikzlibrary{calc}
\pgfdeclarelayer{background}
\pgfdeclarelayer{foreground}
\pgfsetlayers{background,main,foreground}

\usepackage{tikz}  
\usepackage{xcolor}
\definecolor{color1}{HTML}{F0FFF0}
\definecolor{color2}{HTML}{FFEFD5}
\definecolor{color3}{HTML}{FFC0CB}
\definecolor{color4}{HTML}{E8E8E8}

\begin{document}
	
	\title{Three-Body Barrier Dynamics of Double-Alpha Decay in Heavy Nuclei}

	\author{Shulin Tang}
	\affiliation{School of Physics, Nanjing University of Science and Technology, Nanjing 210094, China}
	
	\author{Tao Wan}
	\affiliation{School of Physics, Nanjing University of Science and Technology, Nanjing 210094, China}
	
	\author{Yibin Qian}
	\email{qyibin@njust.edu.cn}
	\affiliation{School of Physics, Nanjing University of Science and Technology, Nanjing 210094, China}
		
	\author{Chong Qi}
	\affiliation{Department of Physics, Royal Institute of Technology (KTH), SE-10691 Stockholm, Sweden}
		
	\author{Ramon A. Wyss}
	\affiliation{Department of Physics, Royal Institute of Technology (KTH), SE-10691 Stockholm, Sweden}
	
	\author{Roberto J. Liotta}
	\affiliation{Department of Physics, Royal Institute of Technology (KTH), SE-10691 Stockholm, Sweden}
	
	\author{Dong Bai}
	\affiliation{College of Science, Hohai University, Nanjing 211100, China}
	
	\author{Bo Zhou}
	\affiliation{Institute of Modern Physics, Fudan University, Shanghai 200433, China}
	
	\author{Zhongzhou Ren}
	\affiliation{School of Physics Science and Engineering, Tongji University, Shanghai 200092, China}
	\newcommand{\RNum}[1]{\uppercase\expandafter{\romannumeral #1\relax}}
	
	\date{\today}
	
\begin{abstract}
The simultaneous emission of two $\alpha$ particles—double-$\alpha$ decay—represents a long-predicted but unobserved mode of nuclear radioactivity. Here we formulate this process as a genuine three-body problem within the hyperspherical coordinate framework and evaluate decay probabilities by numerically solving the corresponding hyperradial Schrödinger equation, combined with large-scale random sampling of the potential parameters; the latter treatment ensures that the present results are more convincing. Inspired by this, we demonstrate that the penetrability ratio between simultaneous and sequential $\alpha$ emission exhibits a strikingly linear dependence on $ZQ_{\alpha\alpha}^{-1/2}$, extending the barrier penetration dynamics into the correlated few-body regime. The nuclei $^{108}$Xe, $^{218}$Ra, $^{224}$Pu, $^{222}$U, $^{216}$Rn, and $^{220}$Th are suggested as the most promising candidates for the observation of double-$\alpha$ decay, with predicted half-lives potentially accessible within present detection limits. Our results provide a unified framework for multi-$\alpha$ decay and open a pathway to probing nuclear clustering and few-body correlations in heavy nuclei.

\end{abstract}


\maketitle

$Introduction.$---Driven by advances in state-of-the-art detection techniques and the availability of intense heavy-ion beams, a variety of exotic decay modes involving the simultaneous emission of two or more identical particles have been systematically investigated—for example, $2\gamma$~\cite{2gamma}, $2n$~\cite{2neutron}, and $2p$~\cite{PhysRevLett.94.232501} as well as the related $2\nu$ and $0\nu$ $2\beta$ decays~\cite{RevModPhys.80.481}. Moreover, rare multiproton emissions ($3p$–$5p$) have also been observed~\cite{3p,4p,5p}, highlighting the expanding landscape of nuclear or radioactive decay modes. These rare processes can provide unique insights into few-body correlations~\cite{few-body} as well as shell effects~\cite{PhysRevLett.110.242502} and the limits of nuclear stability~\cite{Ravl2023}. Beyond their significance for nuclear structure, these processes also have direct implications for astrophysical phenomena, including supernova dynamics~\cite{Supernova}, binary neutron-star mergers~\cite{Barnes_2021, Zhu_2021}, stellar decay rates~\cite{rates}, explosive nucleosynthesis~\cite{Meng-Hua}, and the production of heavy elements~\cite{EPJST2024}. Within this broader context, the simultaneous emission of two $\alpha$ particles emerges as a particularly intriguing case, attracting considerable attention as a novel probe of nuclear structure and multi-particle (cluster) correlations.

In fact, double-$\alpha$ decay in the Hoyle state of $^{12}$C, interpreted as a direct $3\alpha$ breakup, has been investigated in a series of studies~\cite{PhysRevC.49.R1751, PhysRevLett.108.202501, PhysRevLett.113.102501, 12C_1,12C_2}, which established an upper limit for the direct decay branching ratio. By contrast, the simultaneous emission of two $\alpha$ particles from a heavier core, naturally formed as a three-body system, offers a more transparent physical picture: the two $\alpha$ particles are emitted in concert with the residual nucleus acting as a stable spectator. The investigation of 2$\alpha$ decay via the emission of an $^{8}$Be nucleus has been carried out theoretically~\cite{JPL1985, PhysRevC.81.054319}, yet no dedicated experimental searches have so far been conducted for this decay mode. 

Recently, microscopic energy density functionals (EDF) calculations suggest that nuclear deformation governs the mechanism of double-$\alpha$ decay, with two $\alpha$ particles localizing oppositely~\cite{PRL_2a,Zhao}. Phenomenological approaches, including extensions of the unified model for $\alpha$ decay and $\alpha$ capture processes (UMADAC)~\cite{Denisov}, the modified generalized liquid drop model (MGLDM)~\cite{Meg}, the Coulomb and proximity potential model (CPPM)~\cite{San1}, and the preformed cluster model (PCM)~\cite{tbxv-tlbf}, have been employed to estimate double-$\alpha$ decay half-lives across a wide range of nuclei, consistently predicting extremely suppressed probabilities. The experiment conducted by de Marcillac et al.~\cite{deMarcillac2003}, which identified single-$\alpha$ decay in $^{209}$Bi, provided the dataset later used by Tretyak~\cite{Tretyak2021} to set a lower limit of $T_{1/2} > 2.9 \times 10^{20}$ years for double-$\alpha$ emission. Moreover, a dedicated experimental proposal has been put forward at the ISOLDE facility of CERN to probe the double-$\alpha$ decay of $^{224}$Ra~\cite{CERN}. In parallel, experimental efforts on $^{228}$Th and $^{220}$Rn using the FRS Ion Catcher~\cite{FRS} and the GADGET II TPC system~\cite{Meet}, respectively, have demonstrated significantly improved prospects for the experimental observation of this rare decay mode. 

In short, theoretical estimates of double-$\alpha$ decay still exhibit considerable variations, while experimental efforts underscore the urgent need to probe its signatures. In this context, a key quantity for identifying this exotic decay mode is the ratio of its penetrability to that of single-$\alpha$ emission, which provides an essential criterion for experimental discrimination. Complementarily, the corresponding double-$\alpha$ half-life provides a quantitative measure of its observability. One may think that two-proton decay formalisms can be extended to describe the double-$\alpha$ decay process, particularly the simple formalism developed in Ref.~\cite{two-proton}. However, in two-proton decay the mass of the daughter nucleus is much larger than the mass carried by the decaying protons, while in double-$\alpha$ decay the mass of the $\alpha$-particle is large and recoil effects cannot be neglected. But even in two-proton decay a formalism that treats the two-proton decay process within a three-body framework was formulated~\cite{PhysRevC.68.054005}. In what follows, we will describe double-$\alpha$ decay by using that general formalism. That is, we will develop a semiclassical Faddeev-like formalism adapted to double-$\alpha$ decay, and will apply it to determine the most likely candidates that decays through the rare double-$\alpha$ decay channel.

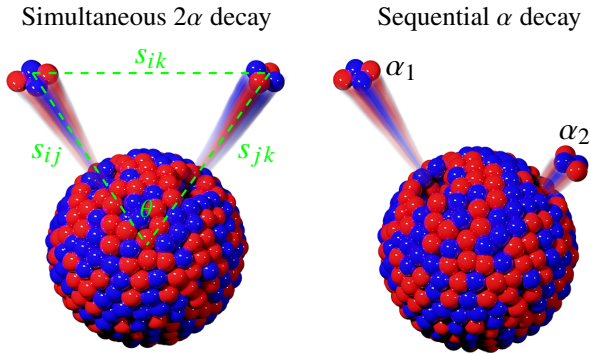
\begin{figure}[htpb]
	\centering
	\begin{tikzpicture}[scale=1.2,>=Stealth]

		\useasboundingbox (-3.6, -1.0) rectangle (3.6, 3.0);
		
		\begin{scope}[on background layer]
			\clip (-3.6, -1.2) rectangle (3.6, 3.3);
			\node[anchor=south west, inner sep=0] at (-3.6,-1.2)
			{\includegraphics[
				width=4.4cm, 
				height=4.7cm, 
				page=1
				]{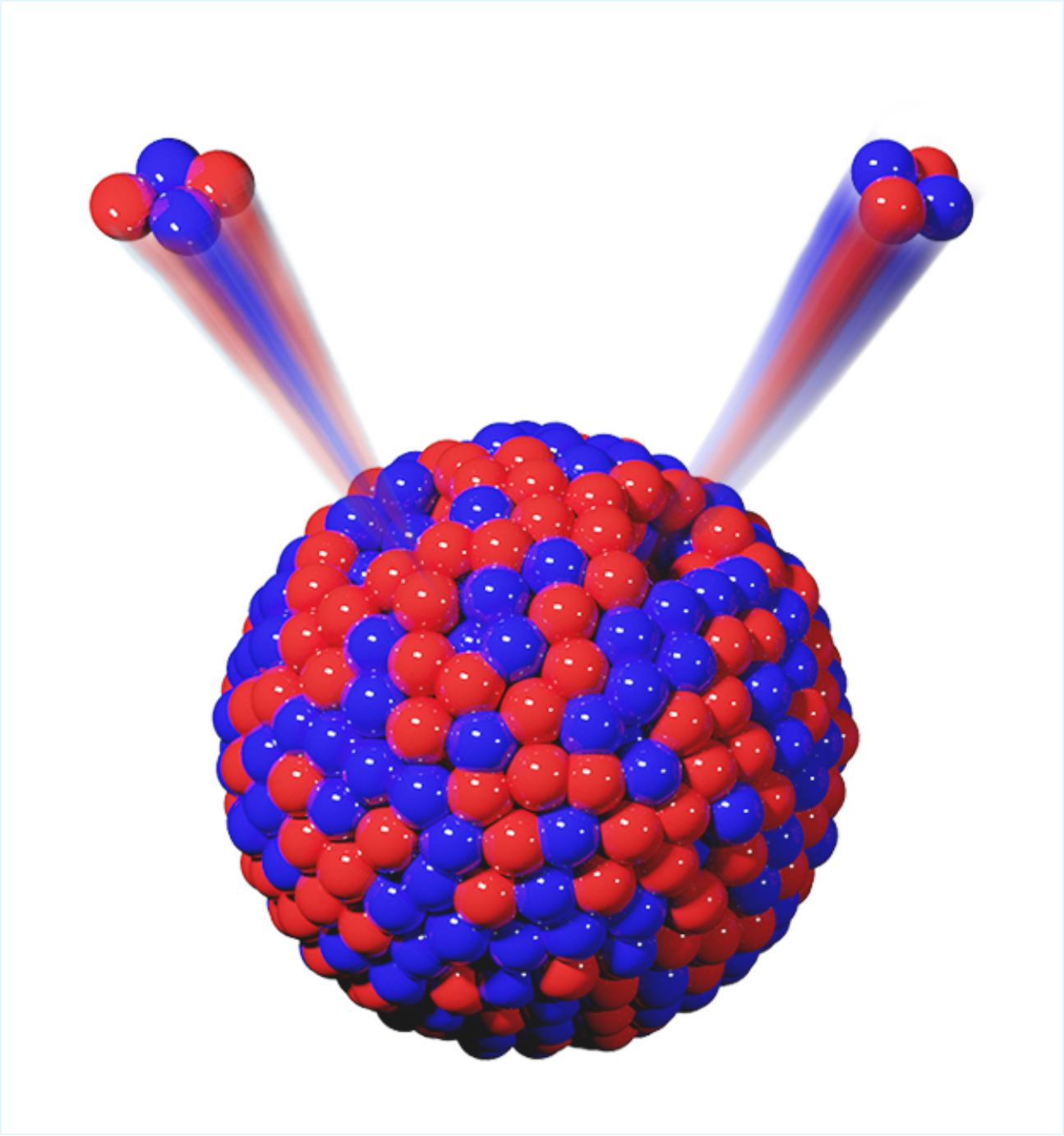}};
			
			\node[anchor=south west, inner sep=0] at (0.1,-1.2)
			{\includegraphics[
				width=4.1cm, 
				height=4.3cm, 
				page=1
				]{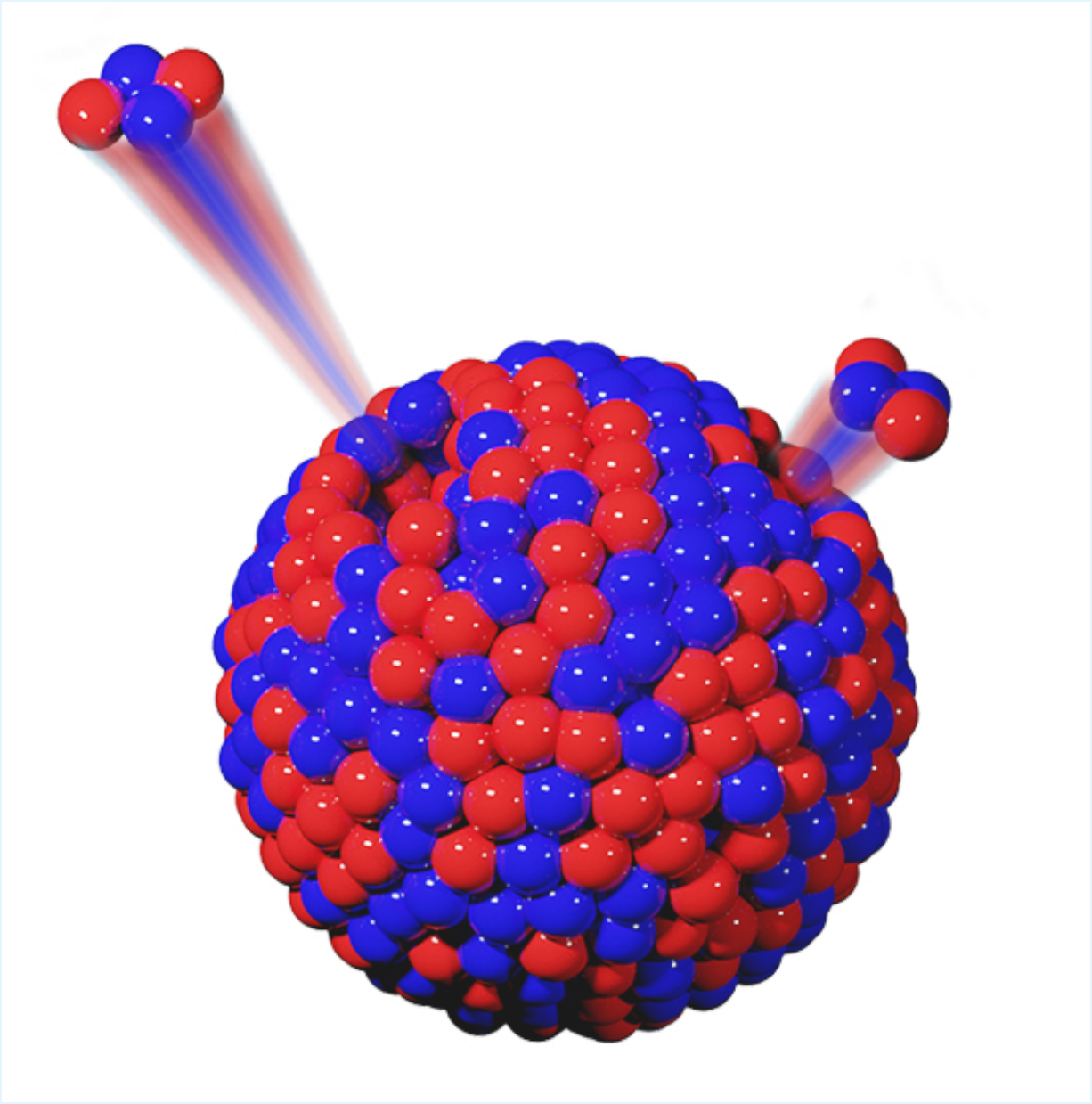}};
		\end{scope}
	
	\definecolor{proton}{RGB}{255, 0, 0}
	\definecolor{neutron}{RGB}{0, 0, 255}
	\definecolor{DarkOrchid}{RGB}{148, 0, 211}
	\definecolor{Green}{RGB}{0, 255, 0}
	
	\coordinate (A1) at (-3.0, 2.1);   
	\coordinate (A2) at (-0.4, 2.1); 
	\coordinate (A3) at (-1.75, 0.2);
	
	\draw[dashed,thick,Green] (A1) -- (A2)
	node[pos=0.6,above left=-2pt,font=\large] {$s_{ik}$};
	\draw[dashed,thick,Green] (A1) -- (A3) node[pos=0.6, above, xshift=-20pt, font=\large] {$s_{ij}$};;
	\draw[dashed,thick,Green] (A2) -- (A3) node[pos=0.6, above, xshift=23pt, font=\large] {$s_{jk}$};
	
	\node[below left=-1pt,font=\large] at (1.3, 2.3) {$\alpha_1$};
	\node[below left=-1pt,font=\large] at (3.2, 1.6) {$\alpha_2$};
	\node[font=\normalsize] at (-1.75, 2.7) {Simultaneous 2$\alpha$ decay};
	\node[font=\normalsize] at (1.9, 2.7) {Sequential $\alpha$ decay};
	\node[font=\normalsize] at (-1.75, 0.6) {$\textcolor{Green}\theta$};
	
	\end{tikzpicture}
	\caption{Schematic diagrams of simultaneous 2$\alpha$ decay (left panel) and sequential $\alpha$ decay (right panel). The right panel actually illustrates two successive $\alpha$ decays of the same nucleus shown in the left panel. In the simultaneous case, the two $\alpha$ particles are emitted without a predetermined order. Note that $s_{ij}, s_{ik}$ and $s_{jk}$ are not the physical distances between the particles, but rather scaled quantities within the hyperspherical formalism.}
	\label{alpha_decay}
\end{figure}

$Theoretical~method.$---Within the framework of hyperspherical coordinate formalism, the dynamics of the three-body system are described in terms of the hyperradius $\rho$~\cite{GARRIDO200527},
\begin{equation}
	\frac{\mathbf{r}^{2}_{ik}}{\rho^{2}} \equiv s_{ik}^{2},
	\quad m M \equiv \sum_{i<k} m_i m_k s_{ik}^2.
\end{equation}
where $\mathbf{r}^{2}_{ik}={(\mathbf{r}_i - \mathbf{r}_k)}^{2}$, $M = \sum_{i} m_{i}$ and $m$ is an arbitrary normalization mass. Assuming that the two simultaneously emitted $\alpha$ particles are emitted from the radioactive nucleus with an arbitrary opening angle $\theta$ (see the left panel of Fig.~\ref{alpha_decay}), the total interaction potential $V(\rho, \theta)$ of the three-body system can be expressed as the sum of the nuclear potential $V_N(\rho, \theta)$, the Coulomb potential $V_C(\rho, \theta)$ and the centrifugal potential:
\begin{equation}
	V(\rho, \theta) = V_N(\rho, \theta) + V_C(\rho, \theta) + \frac{\hbar^{2} (K + 3/2)(K + 5/2)}{2m \rho^{2}},
\end{equation}
where the hypermomentum $K$ is set to 0, as the daughter nuclei formed after double $\alpha$ decay are even-even in their ground states. 

The Coulomb potential in terms of the scaling factors $(s_{ik})$ is expressed as
\begin{equation}
	V_C(\rho, \theta) = \sum_{i<k} \frac{Z_{i} Z_{k} e^{2}}{r_{i k}(\theta)} = \frac{1}{\rho} \sum_{i<k} \frac{Z_{i} Z_{k} e^{2}}{s_{i k}(\theta)}.
\end{equation}
The nuclear potential is taken in the classical Woods-Saxon (WS) form,
\begin{equation}
		V_N(\rho, \theta) = \sum_{i<k} \, \frac{V_{ik}}{1 + \exp\big[(\rho s_{ik}(\theta) - R_{ik}) / a_{ik}\big]}.
	\label{V_nrho}
\end{equation}
Here, $\rho s_{ij}(\theta)$ (or $\rho s_{jk}(\theta)$) denotes the distance between a given $\alpha$ particle and the core of the parent nucleus, and $\rho s_{ik}(\theta)$ corresponds to the distance between the two $\alpha$ particles. The parameters $V$, $R$, and $a$ with subscript $ik$ correspond to the $\alpha$–$\alpha$ interaction potential, and those with subscript $ij$ (or $jk$) refer to the $\alpha$–core interaction potential.

Using the above hyper-radial potential, the radial wave function can be obtained by solving the corresponding hyperradial Schrödinger equation, which is given by
\begin{equation}
\left[ -\frac{\hbar^2}{2\mu_\rho} \frac{d^2}{d\rho^2} + V(\rho, \theta) \right] \varphi_{nlj}(\rho; \theta) = Q_{\alpha\alpha} \varphi_{nlj}(\rho; \theta),
\end{equation}
where $\mu_\rho$ equals $m \times amu$ (or $m \times u$) and $Q_{\alpha\alpha}$ is the double-$\alpha$ decay energy associated with ground-state to ground-state transitions. The quasibound solution $\varphi_{nlj}(\rho; \theta)$ is obtained by imposing the outgoing Coulomb-wave boundary condition
\begin{equation}
\varphi_{nlj}(\rho; \theta) = C(\theta) \left[ F_l(\eta_\rho, k_\rho \rho) + i G_l(\eta_\rho, k_\rho \rho) \right], 
\end{equation}
where wave number $k_\rho = \sqrt{2\mu_\rho Q_{\alpha\alpha} / (\hbar^2)}$ and $C(\theta)$ is the normalization constant. For a certain orientation angle $\theta$ between 2$\alpha$ particles, the corresponding decay width can then be expressed as
\begin{equation}
	\Gamma_{\alpha\alpha}(\theta) = \frac{\hbar^2 k_\rho}{\mu_\rho} \frac{\left| \varphi_{nlj}(\rho_{\text{m}}; \theta) \right|^2}{F_l(\eta_\rho, k_\rho \rho_{\text{m}})^2 + G_l(\eta_\rho, k_\rho \rho_{\text{m}})^2},
\end{equation}
where $\rho_m$ denotes a matching radius in the asymptotic Coulomb region, taken sufficiently large to lie beyond the range of the nuclear interaction. By averaging of $\Gamma_{\alpha\alpha}(\theta)$ in all directions, the penetrability is obtained as
\begin{equation}
	{\Gamma_{\alpha\alpha}}=\int d\theta\,w(\theta)\,\Gamma_{\alpha\alpha}(\theta),
	\qquad
	\int d\theta w(\theta)=1.
	\label{Pavg}
\end{equation}
The angular probability distribution $w(\theta)$, derived from the angular part of the total wavefunction for the present three-body system, is approximated as:
\begin{equation}
	w(\theta) = \left| \mathcal{N} \exp\left(-\frac{(\theta - \pi)^2}{4\sigma^2}\right) \right|^2,
\end{equation}
which favors the back-to-back configuration ($\theta = \pi$). This result is consistent with both the geometric angular factors and the tunneling dynamics established in Refs.~\cite{PRL_2a, Zhao, Denisov}. Further details regarding the derivation are provided in the Supplementary Material.

In an analogous manner, the $\Gamma_\alpha$ of the single-$\alpha$ decay width is obtained from the same Coulomb-matching expression as in the double-$\alpha$ case, with $\varphi_{nlj}(\rho_m; \theta)$ replaced by $\varphi_{nlj}(R)$ and $(\mu_\rho, k_\rho)$ replaced by $(\mu, k)$, where $k$ is defined in terms of the single-$\alpha$ decay energy $Q_\alpha$. In this process, the interaction potential $V(r)$ consists of the nuclear potential $V_N(r) = V / \bigl[1 + \exp((r - R^{\prime}) / a)\bigr]$, the Coulomb potential $V_C(r)$, and the centrifugal potential~\cite{XuChang}. The nuclear radius parameter $R^{\prime}$ in the sequential decay scenario is determined separately by the Wildermuth condition~\cite{XuChang, wildermuth}, where $G = 2n + \ell = \sum_{i=1}^{4} g_i$. In this context, the nucleus remaining after the simultaneous emission of two $\alpha$ particles is equivalent to that obtained after the emission of $\alpha_2$ from the intermediate nucleus in the sequential-decay mode (see the right panel of Fig.~\ref{alpha_decay}). Accordingly, the radius parameter $R_{ij}$ (or $R_{jk}$) in Eq.~(\ref{V_nrho}) is identified with $R'$, representing the nuclear radius in the sequential-decay picture at the emission of the second $\alpha$ particle. 

A direct comparison between the decay widths of double-$\alpha$ and single-$\alpha$ decay offers not only a quantitative measure of the relative likelihood of this exotic mode but also a practical criterion for identifying the most promising nuclei for experimental searches, thereby motivating the definition of the branching ratio (BR) as
\begin{equation}
	\mathrm{BR} = \text{log}_{10}(\frac{\Gamma_{\alpha\alpha}}{\Gamma_\alpha}).
\end{equation}
The half-life of double-$\alpha$ decay is defined as
\begin{equation}
	T_{\alpha\alpha} = \frac{\hbar \ln 2}{P_{0}\Gamma_{\alpha\alpha}},
\end{equation}
where $P_0$ is the preformation factor, which is evaluated using the cluster formation model (CFM)~\cite{Deng_2015, PhysRevC.93.044326}.

\begin{figure}[htbp]
	\centering
	\begin{tikzpicture}
	
		\newlength{\procW}
		\setlength{\procW}{4.65cm}
		
		\tikzstyle{process} = [draw, rectangle, minimum width=5cm, minimum height=0.2cm, align=flush left, thick];
		\tikzstyle{process1} = [draw, rectangle, minimum width=5cm, minimum height=0.2cm, align=flush left, thick, fill=color1, rounded corners];
		\tikzstyle{process2} = [draw, rectangle, minimum width=5cm, minimum height=0.1cm, align=flush left, thick, fill=color2, rounded corners];
		\tikzstyle{process3} = [draw, rectangle, minimum width=5cm, minimum height=0.2cm, align=flush left, thick, fill=color3, rounded corners];
		\tikzstyle{processM} = [rectangle, minimum width=5cm, minimum height=0.2cm, align=flush left, rounded corners];
		\tikzstyle{processBG} = [draw, rectangle, minimum width=5cm, minimum height=5.5cm, very thin, fill=color4];
		\tikzstyle{decision} = [draw, diamond, minimum width=3cm, minimum height=0.2cm, aspect=2, align=center, thick, fill=color3];  
		\tikzstyle{connector} = [->, >=latex, very thick];  
		\tikzstyle{line} = [-, >=latex, very thick]; 

		\path[use as bounding box] (-4.4,-2.6) rectangle (4.4,2.6);
		
		\node (origin) at (0,0) {};
		
		\node (A) [process1, minimum width=3cm, left=1cm of origin, yshift=1.8cm,
		align=center, font=\fontsize{8pt}{8.5pt}\selectfont] 
		{
			INPUT PARAMETERS\\
			Para-$\alpha_1$: $Q_{\alpha1}$, $G_1$\\
			Para-$\alpha_2$: $Q_{\alpha2}$, $G_2$\\
			Para-$\alpha\alpha$: $Q_{\alpha\alpha}$
		};
	
		\node (BG) [processBG, minimum width=4.9cm, right=-0.9cm of origin]{};
		
		\node (B) [process2, minimum width=4.5cm, right=-0.75cm of origin, yshift=1.8cm,
		align=center, font=\fontsize{8pt}{8.5pt}\selectfont] 
		{
			Randomly sample WS parameters with \\
			120 $< V <$ 220 and 0.3 $< a <$ 0.7
		};
	
		\node (M) [processM, minimum width=\procW, above=0.05cm of B,
		align=center, font=\fontsize{9pt}{8.5pt}\selectfont] 
		{
			Large-scale Random Sampling Loop
		};
	
		\node (C) [process2, minimum width=\procW, below=0.25cm of B,
		align=center, font=\fontsize{8pt}{8.5pt}\selectfont] 
		{
			Calculate sequential $\alpha$ decay width $\Gamma_\alpha$ \\
			using Para-$\alpha_1$, Para-$\alpha_2$. Extract radius \\
			parameter $R^{\prime}$ from sequential channel.
		};
	
		\node (D) [process2, minimum width=\procW, below=0.25cm of C,
		align=left, font=\fontsize{8pt}{8.5pt}\selectfont] 
		{
			Calculate simultaneous decay width  \\ 
			$\Gamma_{\alpha\alpha}$ using Para-$\alpha\alpha$ and extracted $R^{\prime}$, \\
			and then compute BR and $T_{\alpha\alpha}$.
		};
	
		\node (E) [decision, minimum width=2cm, below=0.25cm of D,
		align=left, font=\fontsize{8pt}{8.5pt}\selectfont] 
		{ $i$ $>$ 5,000?};
		
		\node (F) [process1, minimum width=3cm, below=2.5cm of A,
		align=center, font=\fontsize{8pt}{8.5pt}\selectfont] 
		{
           STATISTICAL ANALYSIS:\\
           Final values and  \\
           uncertainties of BR and  \\
           $T_{\alpha\alpha}$ from 5,000 samples.
		};
		
		\coordinate (A1) at ($(A.east)$);
		\coordinate (A2) at ($(A.east) + (+0.50cm, 0cm)$);
		
		\coordinate (B1) at ($(B.south)$);
		\coordinate (B2) at ($(B.south) + (0cm, -0.26cm)$);
		
		\coordinate (C1) at ($(C.south)$);
		\coordinate (C2) at ($(C.south) + (0cm, -0.26cm)$);
		
		\coordinate (D1) at ($(D.south)$);
		\coordinate (D2) at ($(D.south) + (0cm, -0.26cm)$);
		
		\coordinate (E1) at ($(E.west)$);
		\coordinate (E2) at ($(E.west) + (-1.45cm, 0cm)$);
		
		\coordinate (E3) at ($(E.east)$);
		\coordinate (E4) at ($(E.east) + (+1.45cm, 0cm)$);
		
		\coordinate (E41) at ($(E.east) + (+1.45cm, -0.021cm)$);
		\coordinate (B3) at ($(B.east)$);
		
		\draw [connector] (A1) -- (A2);
		\draw [connector] (B1) -- (B2);
		\draw [connector] (C1) -- (C2);
		\draw [connector] (D1) -- (D2);
		\draw [connector] (E1) -- node[above, font=\bfseries, yshift=-0.05cm] {Yes} (E2);
		\draw [line] (E3) -- node[above, font=\bfseries, yshift=-0.05cm] {No} (E4);
		\draw [connector] (E41) |- (B3);

	\end{tikzpicture}
	\caption{Flowchart for the calculation of half-lives of double-$\alpha$ decay and the decay width branching ratio $\mathrm{BR} = \Gamma_{\alpha\alpha}/\Gamma_\alpha$. The terms Para-$\alpha_1$, Para-$\alpha_2$ and Para-$\alpha\alpha$ denote the parameters associated with the sequential emissions of $\alpha_1$ and $\alpha_2$ and simultaneous $2\alpha$ emission, respectively.}
	\label{flowchart}
\end{figure}
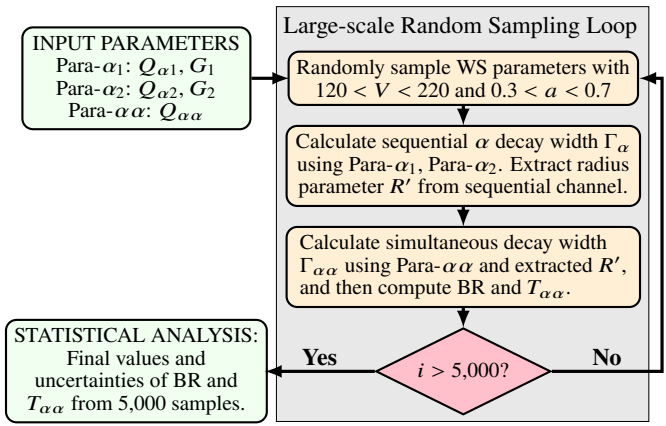

Because the WS interaction parameters entering Eq.~(\ref{V_nrho}) are not uniquely constrained for the present three-body problem, we treat them as nuisance parameters and propagate their associated uncertainties in a statistically controlled manner. Specifically, a large ensemble of parameter sets is sampled over physically reasonable ranges, and the requirement of a consistent description of the sequential single-$\alpha$ decay is imposed for each set within the same framework. This procedure anchors the effective interaction to the well-established $\alpha$-decay channel and thereby enables a prediction of the double-$\alpha$ width $\Gamma_{\alpha\alpha}$ without further parameter adjustment. Importantly, the inferred BR and $T_{\alpha\alpha}$ are found to be robust against reasonable variations of the interaction. The complete workflow is summarized in Fig.~\ref{flowchart}:
\begin{enumerate} \renewcommand{\labelenumi}{(\textit{\roman{enumi}})}
	\item Specify the simultaneous $2\alpha$ decay energy $Q_{\alpha\alpha}$, as well as the sequential $\alpha$-decay energies $Q_{\alpha1}$ and $Q_{\alpha2}$, together with the corresponding global quantum numbers $G_1$ and $G_2$. 
	\item Randomly sample the WS parameters over a sufficiently wide range. For each sampled set, calculate $\Gamma_\alpha$ and extract the radius parameter $R^{\prime}$ from the sequential-decay channel.
	\item Using the same set of WS parameters, compute $\Gamma_{\alpha\alpha}$ with $Q_{\alpha\alpha}$ and the extracted $R^{\prime}$, and then evaluate BR and $T_{\alpha\alpha}$. The final values and associated uncertainties are obtained from a statistical analysis of the results after 5,000 random samples of the WS parameters.
\end{enumerate}

$Results~and~discussion.$---The double-$\alpha$ decay of the most promising candidate nuclei, as indicated by both theoretical predictions and experimental considerations, have been systematically investigated within a semiclassical Faddeev-like formalism in hyperspherical coordinates by solving the hyperradial Schrödinger equation. According to the above procedure, 5,000 random samples of the interaction-potential parameters are generated to ensure a consistent description of both single- and double-$\alpha$ decay (see Fig.~\ref{flowchart}). To enable a direct comparison of the two decay channels, the ratio $\mathrm{BR} = \Gamma_{\alpha\alpha}/\Gamma_\alpha$ is determined for each sampled parameter set, along with the corresponding double-$\alpha$ half-life $T_{\alpha\alpha}$. Ensemble averages and uncertainties for $\mathrm{BR}$ and $T_{\alpha\alpha}$ are then extracted from this statistical set, providing a robust quantitative assessment of the decay properties.

Notably, the half-lives of sequential $\alpha$ decay calculated with the above random-sampling procedure, where the $P_0$ is consistently evaluated within the CFM, agree excellently with experimental data. This consistency provides strong justification for applying the same sampling strategy to double-$\alpha$ decay. Moreover, the error-bars in the calculated single-$\alpha$ decay half-lives, arising from the strategy of random sampling, remain within roughly 1.2 orders of magnitude for almost all nuclei, further highlighting the robustness of the present framework. 

\begin{figure}[htpb]
	\setlength{\abovecaptionskip}{0pt}
	\centering
	{
		\includegraphics[width=0.5\textwidth,height=0.20\textheight]{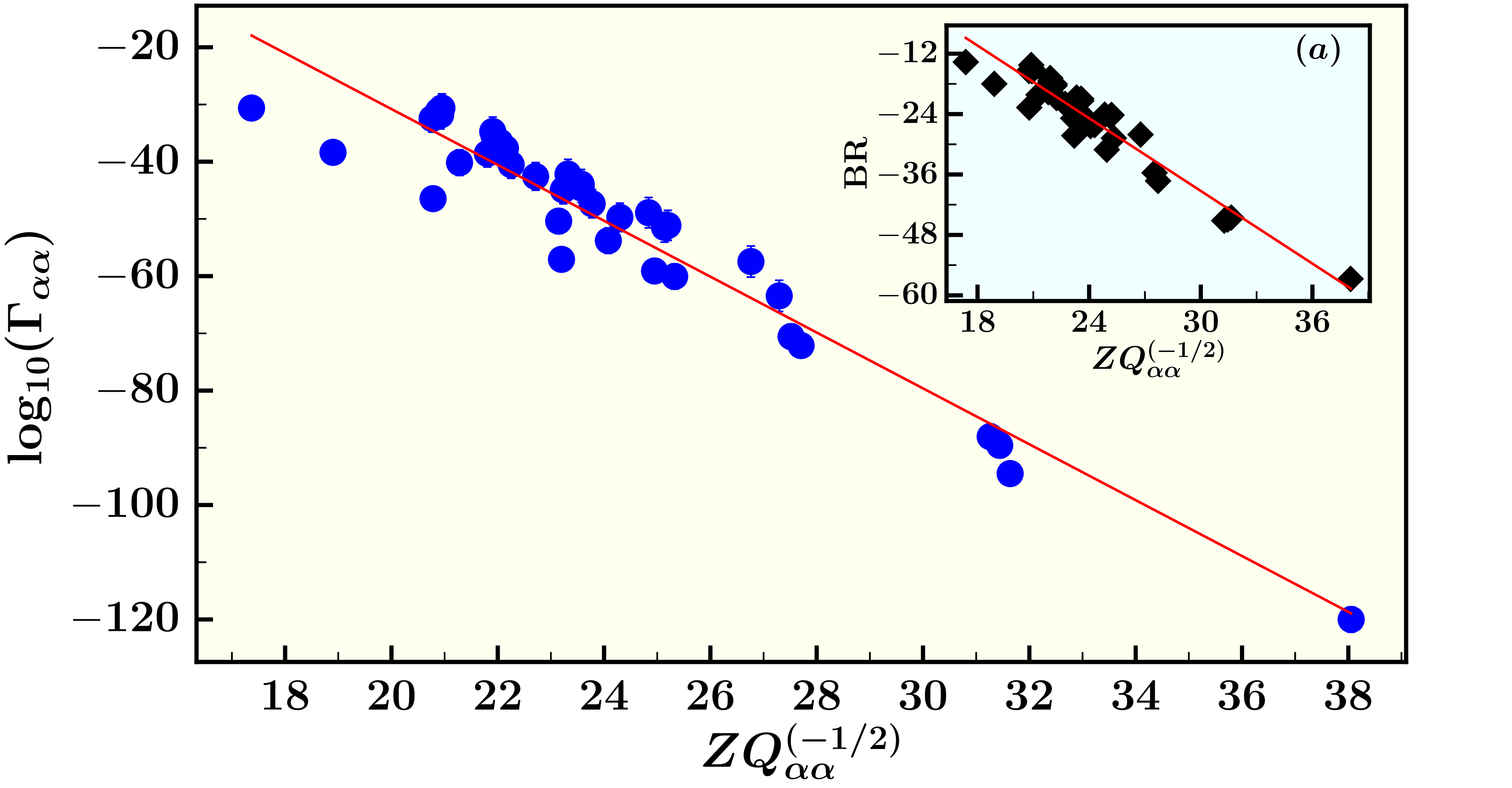}
	}
	\vspace*{0pt}
	\caption{The main panel shows the logarithm of the penetrability ratio $\mathrm{log_{10}(\Gamma_{\alpha\alpha})}$ as a function of $ZQ^{-1/2}$, with the inset (a) displays BR as a function of the same variable. The linear fit in main panel yields a Pearson correlation coefficient of $0.961$, indicating that the half-lives of double-$\alpha$ decay approximately follow a Geiger–Nuttall type relation.}
	\label{ratio}
\end{figure}

As illustrated in the main panel of Fig.~\ref{ratio}, the calculated decay width ratios $\mathrm{log_{10}(\Gamma_{\alpha\alpha})}$ exhibit a strong linear correlation with $ZQ^{-1/2}$ (with a Pearson correlation coefficient $r_p$ = $0.963$), reminiscent of Geiger–Nuttall law for single-$\alpha$ decay. This systematic trend indicates that double-$\alpha$ decay is likewise governed by barrier penetration dynamics, offering a practical scaling relation for identifying promising candidate nuclei. As shown in the inset (a) of Fig.~\ref{ratio}, it is interested that the BR versus $ZQ^{-1/2}$ displays a similarly pronounced linear behavior ($r_p$ = $0.959$), suggesting that the competition between simultaneous and sequential emission is also largely controlled by the same Coulomb-dominated barrier. 

\begin{figure}[htpb]
	\setlength{\abovecaptionskip}{0pt}
	\centering
	{
		\includegraphics[width=0.5\textwidth,height=0.20\textheight]{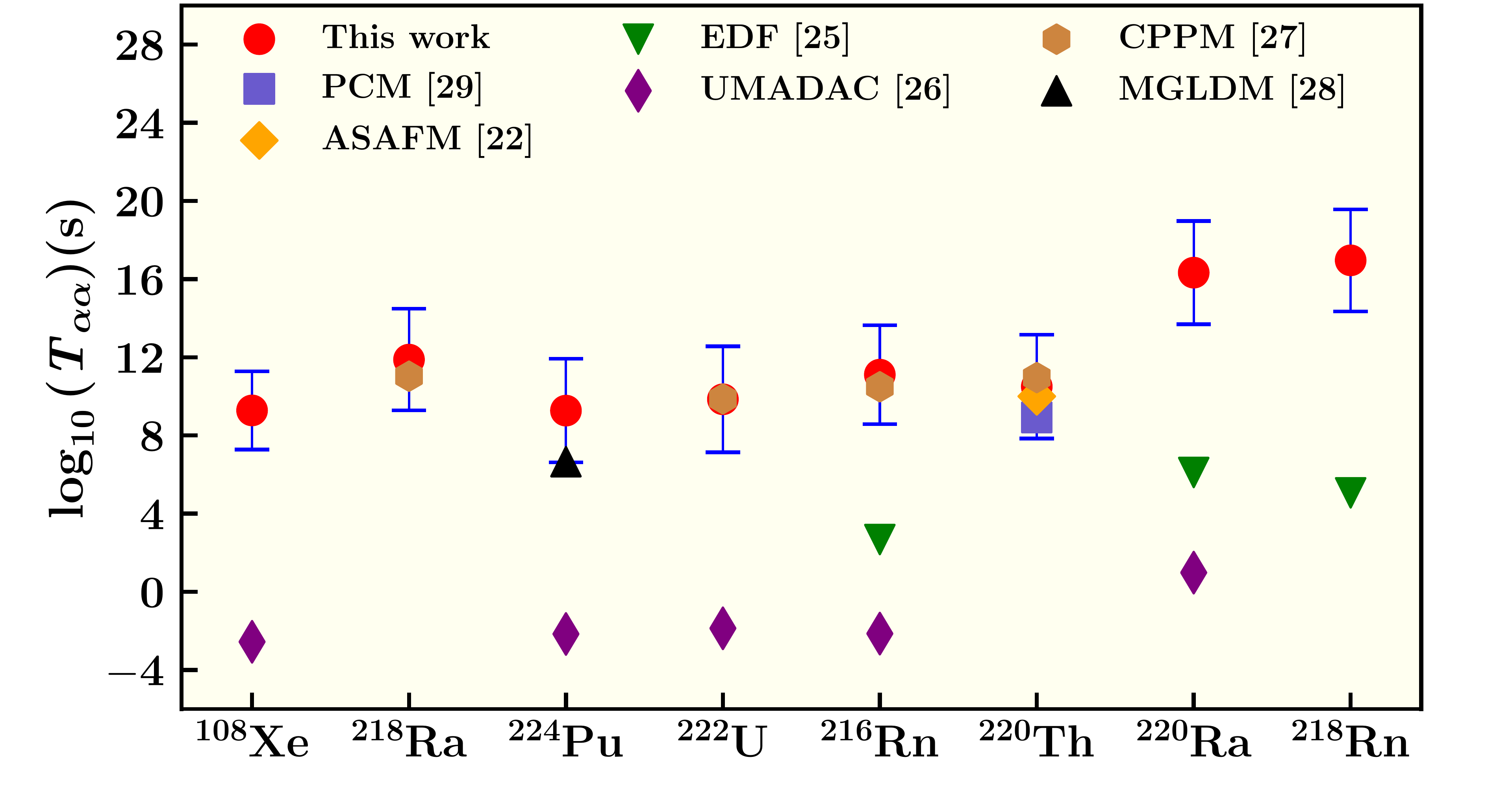}
	}
	\vspace*{0pt}
	\caption{Logarithm of the simultaneous 2$\alpha$ decay half-life, $\log_{10}(T_{\alpha\alpha})$, for the promising candidate nuclei. Red circles denote the present prediction using the Faddeev-like formalism, with vertical error bars reflecting the statistical uncertainties obtained from the parameter-sampling procedure. Other symbols correspond to theoretical estimates reported in Refs.~\cite{JPL1985, Denisov, Zhao, San1, Meg, tbxv-tlbf}.}
	\label{life}
\end{figure}

From a physical perspective, the simultaneous emission of two $\alpha$ particles becomes more favorable as the ratio $\mathrm{BR}$ increases. Moreover, the dispersion in the branching ratio, quantified by one standard deviation from the statistical sampling, does not exceed 1.4 orders of magnitude for all nuclei, underscoring both the robustness and the reliability of the present framework. In the present work, $^{108}$Xe, $^{218}$Ra, $^{224}$Pu, $^{222}$U, $^{216}$Rn, and $^{220}$Th are identified as the most favorable candidates for double-$\alpha$ emission, with probability ratios of $10^{-13.67}$, $10^{-14.28}$, $10^{-15.00}$, $10^{-15.38}$, $10^{-15.42}$ and $10^{-15.46}$, respectively. By comparison, the $^{14}$C radioactivity of $^{223}$Ra exhibits the largest experimentally measured cluster branching ratio, $b \approx 10^{-8.9}$ relative to $\alpha$ decay~\cite{HPR}. The double-$\alpha$ half-lives of these most promising candidates are shown in Fig.~\ref{life}. The red circles denote $\log_{10}\bigl(T_{\alpha\alpha}/\mathrm{s}\bigr)$ for $^{108}$Xe, $^{218}$Ra, $^{224}$Pu, $^{222}$U, $^{216}$Rn, and $^{220}$Th, with central values of $9.28$, $11.88$, $9.27$, $9.85$, $11.11$, and $10.50$, respectively. The vertical error bars indicate the $1\sigma$ uncertainties extracted from the WS parameter-sampling procedure. Encouragingly, the predicted double-$\alpha$ half-lives are comparable to the shortest cluster decay observed so far, the $^{14}$C emission from $^{222}$Ra with $T_{1/2}=10^{11.01}$ s~\cite{HPR}, highlighting the potential experimental accessibility of double-$\alpha$ decay in these nuclei. 

By contrast, the single-$\alpha$ half-lives of these promising candidates are much shorter than those of the other nuclei considered in this work, with $\log_{10}(T_{\alpha}/\mathrm{s})$ lying in the range from -6 to -4. Moreover, the corresponding single-$\alpha$ half-lives of their daughter nuclei fall in the range -9 to -6, indicating that both the parent and daughter nuclei are strongly $\alpha$-unstable and thus constitute favorable candidates for double-$\alpha$ emission. Previous studies have suggested that double-$\alpha$ decay is enhanced when the two-$\alpha$ daughter lies closer to $^{208}$Pb. Consistent with this picture, our calculations show that, among such daughters, configurations near the neutron shell closure N=126 exhibit a higher double-$\alpha$ likelihood and shorter $T_{\alpha\alpha}$ than cases associated primarily with the proton closure Z=82. This points to the enhanced binding at N=126 as a common structural driving force shaping both double- and single-$\alpha$ decay systematics, in agreement with the experimental trends of single-$\alpha$ half-lives.

As one would expect, $^{212}$Po, corresponding to the $^{208}$Pb+$\alpha$ configuration, shows a strongly suppressed double-$\alpha$ decay, with $\log_{10}(\mathrm{BR})=-49.07$ and $\log_{10}(T_{\alpha\alpha} / s)=42.11$. In contrast, $^{216}$Rn, associated with the more symmetric $^{208}$Pb+$2\alpha$ configuration, exhibits a much enhanced likelihood for this mode, with the present calculation yielding a significantly shorter half-life of $\log_{10}(T_{\alpha\alpha} / s)=11.11$. Notably, $^{216}$Rn has also been highlighted in Ref.~\cite{Zhao} as a prime candidate, in good agreement with the present findings. This result is further supported by systematics of Heavy-Particle Radioactivity (HPR)~\cite{HPR}, where the most probable clusters emitted from $^{222}$Ra and $^{288}$Fl are $^{14}$C and $^{80}$Ge, respectively, both leading to the $^{208}$Pb daughter nucleus. The consistency between these decay pathways points to the influence of the $^{208}$Pb shell closure in favoring both cluster and exotic decay modes. Furthermore, the double-$\alpha$ decay of $^{108}$Xe is particularly appealing because it populates the doubly magic nucleus $^{100}$Sn. In addition, experimentally relevant candidates—$^{220}$Rn~\cite{Meet}, $^{224}$Ra~\cite{CERN}, and $^{228}$Th~\cite{FRS}—have been identified, with $\mathrm{BR}$ values of $-21.37$, $-24.18$, and $-28.08$, corresponding to $\log_{10}(T_{\alpha\alpha} / s)$ = 23.24, 29.80, and 36.11, respectively.

More than a dozen candidate nuclei are predicted to exhibit $\log_{10}(\mathrm{BR})>-20$, typically associated with comparatively short simultaneous 2$\alpha$ decay and sequential $\alpha$ half-lives. Although most of the predicted candidates lie in the heavy region ($Z=86$–94), $^{110}$Xe and $^{160}$W also exceed this threshold and provide complementary light-mass test cases.

Previous theoretical studies of double-$\alpha$ emission have employed energy-density-functional and cluster/phenomenological frameworks to estimate branching ratios. Our work differs fundamentally by treating the process as a genuine three-body problem in hyperspherical coordinates and by propagating Woods–Saxon parameter uncertainties through large-scale numerical solutions of the Schrödinger equation; this approach reveals a robust, nearly linear scaling of $\log_{10}(\Gamma_{\alpha\alpha}/\Gamma_{\alpha})$ with $ZQ_{\alpha\alpha}^{-1/2}$, a relation not reported in earlier phenomenological surveys. The process of double-$\alpha$ emission is conceptually analogous to two-proton decay, in which simultaneous emission of two identical nucleons from a proton-rich nucleus can proceed through distinct mechanisms—sequential decay, direct three-body (true 2p) decay, and correlated di-proton (s-wave anti-symmetric fermion pair) emission~\cite{2p, PhysRevC.82.014615, PhysRevLett.120.212502, PhysRevC.106.034602, ZHANG2023, PFUTZNER2023104050}. In the present context, the heavy-nucleus analogues are sequential $\alpha$ emission, simultaneous (true) double-$\alpha$ decay, and emission of a correlated $^{8}$Be cluster. The hyperradial Schrödinger formulation adopted here corresponds to the ``true" three-body tunneling limit, where both $\alpha$ particles penetrate the barrier coherently. This analogy underscores the universal role of few-body correlations in barrier-penetration phenomena across the nuclear chart, linking proton-dripline dynamics with cluster emission in heavy systems.
	
In summary, the potential candidates for double-$\alpha$ decay have been systematically investigated as a three-body process within the hyperspherical coordinate formalism. By employing large-scale random sampling of Woods–Saxon parameters, we established a consistent description of both single- and double-$\alpha$ decay. A nearly linear scaling of the penetrability ratio with $ZQ_{\alpha\alpha}^{-1/2}$ was revealed, reminiscent of the Geiger–Nuttall law, underscoring the barrier-penetration nature of this exotic decay. The analysis highlights several promising candidates, most notably $^{108}$Xe, $^{218}$Ra, $^{224}$Pu, $^{222}$U, $^{216}$Rn, and $^{220}$Th, whose predicted half-lives fall within a range that could be accessible experimentally. These results provide both theoretical benchmarks and practical guidance for future searches. Continued progress requires more sophisticated treatments of deformation and dynamic correlations, supported by next-generation detection capabilities. The eventual observation of double-$\alpha$ decay would provide a unique probe of clustering and its impact on nuclear structure and astrophysical processes.

\section*{ACKNOWLEDGMENTS}
This work is supported by the National Natural Science Foundation of China (Grants No. 12075121, and No. 12035011), and by the Natural Science Foundation of Jiangsu Province (Grant No. BK20190067).

\bibliography{ref2alpha}
	
\end{document}